\documentclass[10pt,journal,twoside,final]{IEEEtran}

\IEEEoverridecommandlockouts

\usepackage[hyphens]{url}

\usepackage{graphicx} 
\usepackage{bigstrut}
\usepackage{multirow}
\ifCLASSOPTIONcompsoc
\usepackage[caption=false,font=normalsize,labelfont=sf,textfont=sf]{subfig}
\else
\usepackage[caption=false,font=footnotesize]{subfig}
\fi
\usepackage{setspace}

\usepackage{color}
\definecolor{c}{rgb}{0,0,0}
\definecolor{c2}{rgb}{1,0,0}
\definecolor{rev}{rgb}{1,0,0}
\usepackage{bigstrut}
\usepackage{cite,amsfonts}
\usepackage[cmex10]{amsmath}
\usepackage[subnum]{cases}
\usepackage{array}
\newcolumntype{L}[1]{>{\raggedright\let\newline\\\arraybackslash\hspace{0pt}}m{#1}}
\newcolumntype{C}[1]{>{\centering\let\newline\\\arraybackslash\hspace{0pt}}m{#1}}
\newcolumntype{R}[1]{>{\raggedleft\let\newline\\\arraybackslash\hspace{0pt}}m{#1}}

\usepackage{algorithm}
\usepackage{algpseudocode}
\algnewcommand\algorithmicinput{\textbf{INPUT:}}
\algnewcommand\INPUT{\item[\algorithmicinput]}
\algnewcommand{\algorithmicoutput}{\textbf{OUTPUT:}}
\algnewcommand\OUTPUT{\item[\algorithmicoutput]}
\algrenewcommand{\algorithmiccomment}[1]{\hskip3em$\%$ #1}

\begin{document}

\title{Securing Wireless Communications of the Internet of Things from the Physical Layer, An Overview}
\author{
Junqing~Zhang,
Trung~Q.~Duong,~\IEEEmembership{Senior~Member,~IEEE,}
Roger~Woods,~\IEEEmembership{Senior~Member,~IEEE,} 
and
Alan~Marshall,~\IEEEmembership{Senior~Member,~IEEE}
\thanks{This work was supported by the Royal Society Research Grant under Grant ID RG160302. 
}
\thanks{J. Zhang, T. Q. Duong and R. Woods are with School of Electronics, Electrical Engineering and Computer Science, Queen's University Belfast, Belfast, BT9 5AG, U.K. (email: jzhang20@qub.ac.uk; trung.q.duong@qub.ac.uk; r.woods@qub.ac.uk)}
\thanks{A. Marshall is with Department of Electrical Engineering and Electronics,  University of Liverpool, Liverpool, L69 3GJ, U.K. (email: Alan.Marshall@liverpool.ac.uk)}

}

\maketitle

\setcounter{page}{1} \thispagestyle{plain}

\begin{abstract}
The security of the Internet of Things (IoT) is receiving considerable interest as the low power constraints and complexity features of many IoT devices are limiting the use of conventional cryptographic techniques. This article provides an overview of recent research efforts on alternative approaches for securing IoT wireless communications at the physical layer, specifically the key topics of key generation and physical layer encryption. These schemes can be implemented and are lightweight, and thus offer practical solutions for providing effective IoT wireless security. Future research to make IoT-based physical layer security more robust and pervasive is also covered.
\end{abstract}



\section{Introduction}\label{section:intro}
The Internet of Things (IoT) aims to allow ubiquitous connections between things with computing, communication, and sensing ability. IoT applications include smart cities, smart traffic, healthcare, smart home, industrial monitoring, and environment monitoring, etc. \cite{atzori2010internet,al2015internet}, which have revolutionized every aspect of our life. On the other hand, many open research problems still remain to allow this technology to become widely available as proposed~\cite{stankovic2014research}. For example, IoT security and privacy remains a major concern as indicated by the UK IoT government report \cite{walport2014iot} and agreed by many experts~\cite{mckinsey2015iot}. To this end, research into effective IoT security remains a key objective as indicated by major sponsors such as the US National Science Foundation \cite{nsf}, the European Horizon 2020 research programs \cite{ierc}, and the UK's Engineering and Physical Sciences Research Council \cite{epsrc2016iothub}.

The number of connected devices has already exceeded the world's population and is increasing exponentially. It is predicted by numerous sources that IoT devices will number 10 billion by 2020~\cite{nordrum2016iot}. For example, Cisco estimated there would be 6.58 connected devices per person by 2020, i.e., about 50 billion devices in total~\cite{cisco2011iot}.
With the huge amount of IoT devices, wireless communication is preferred as it allows easy installation and provides ubiquitous connection. Wireless air interfaces involved in the IoT include IEEE 802.15.4 (Zigbee), Bluetooth Low Energy (BLE), IEEE 802.11 Wi-Fi, LoRaWAN, cellular connections, ultrawide band, near field communication (NFC), radio-frequency identification (RFID), to name but a few.

While we are enjoying the benefits that wireless connection has brought, its broadcast nature makes the transmission vulnerable to passive eavesdropping and active jamming. Thus, there is a clear need to protect the data on-the-fly in the IoT as it will generally contain sensitive, private or confidential information. For example, in healthcare applications, the sensor nodes collect patients' health information such as heart rate and blood pressure. This information is private and highly confidential, and hence a secure transmission channel is required. However, IoT systems are far from safe and many vulnerabilities exist~\cite{grau2016iot}. For example, HP reported that 70\% of devices did not encrypt their communications~\cite{hp2015iot}.

\begin{figure*}[!t]
	\centering
		\includegraphics[width=6in]{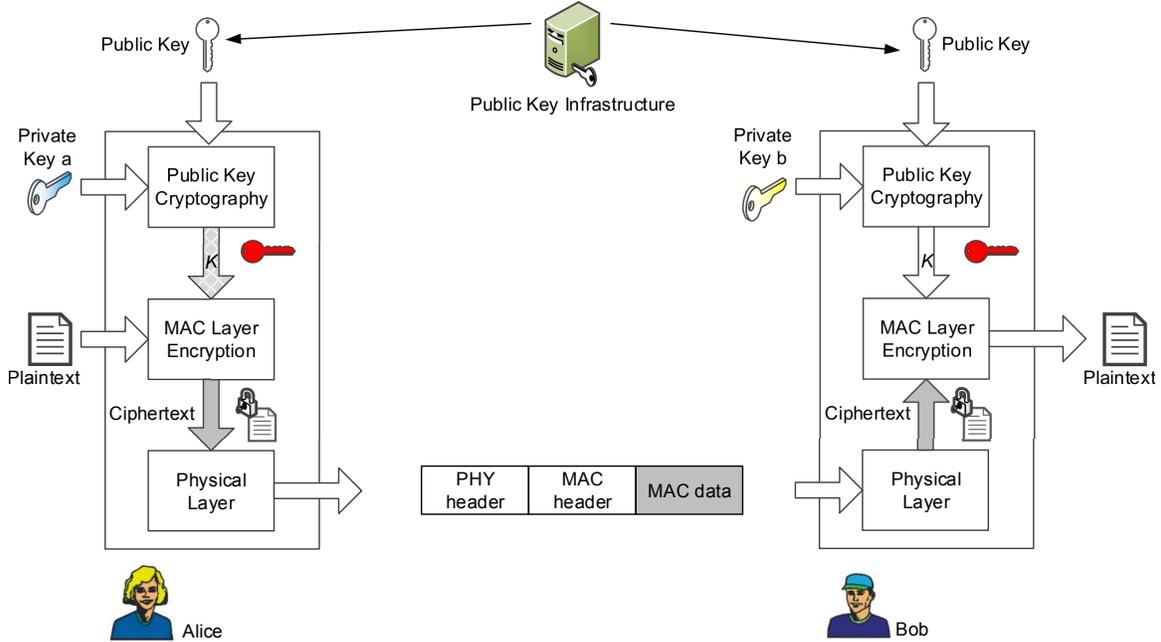}
	\caption{Classic cryptosystem with MAC layer encryption as an example. The gray blocks represent the encrypted data.}
		\label{fig:EncyptionSystem}
\end{figure*}

The security countermeasures are mainly categorized into computational security and information-theoretic security~\cite{zou2016survey}. The former has been the main approach in protecting the communication systems where cryptographic algorithms and protocols are deployed at the upper layers of the protocol stack~\cite{granjal2015security}. For example, the transport layer security (TLS) is a well-known protocol to protect the transport link~\cite{rfc5246tls} while Wi-Fi protected access (WPA) is designed to secure the media access control (MAC) layer\footnote{IEEE 802 splits the open systems interconnection (OSI) data link layer into MAC sublayer and logical link control (LLC) sublayer.} in the IEEE 802.11 systems~\cite{ieee2004security}.
A classical cryptosystem comprises public key cryptography (PKC) for key distribution\footnote{Public key cryptography can also be used for encryption/decryption and digital signature, which are not discussed in this article.} and symmetric encryption for data protection, as shown in Figure~\ref{fig:EncyptionSystem}, where Alice and Bob are the legitimate users wishing to communicate securely. PKC security relies on exploiting the computational hardness of mathematical problems, such as discrete logarithm, and distributes the same session key to Alice and Bob.
Symmetric encryption usually occurs in the upper layers, i.e., data link/MAC layer and above, allowing encryption of plaintext with the common session key shared between users using PKC.

Whilst classical cryptosystems have protected conventional wireless systems, there are challenges in applying these approaches in IoT. IoT devices range from well-resourced smartphones to low cost, low energy and lightweight computing embedded devices. Many low cost IoT devices cannot afford the additional silicon area, power consumption, and code space needed to perform the expensive mathematical calculations of cryptographic methodologies~\cite{trappe2015low}. In addition, IoT applications may work in a device-to-device communication mode where there is no secured public key infrastructure (PKI) for the distribution of public keys. Finally, with the development of quantum computing, the concept of PKC will be fundamentally challenged~\cite{cheng2017securing}.

Conventional upper layer-based cryptography also leaves the transmission vulnerable to many passive and active attacks. For example, the MAC header is sent in plaintext and attackers can perform traffic analysis by observing the MAC header. In addition, the physical packet header is also sent in plaintext and can reveal side-channel information (SCI) such as data rate, packet length, mapping schemes, etc.~\cite{rahbari2015secrecy}. Eavesdroppers can perform various attacks based on the observed SCI, such as analysis of users' activities and selective jamming.

\begin{figure*}[!t]
	\centering
		\includegraphics[width=4in]{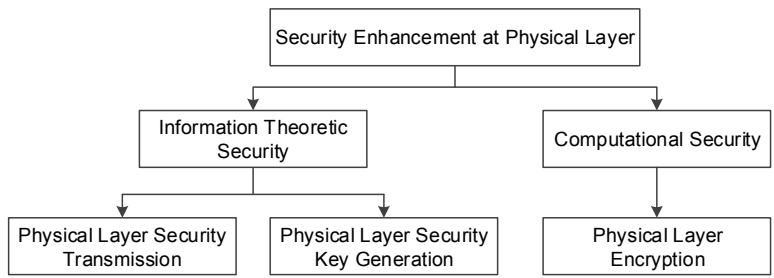}
	\caption{Taxonomy of security enhancement techniques at physical layer}
		\label{fig:taxonomy}
\end{figure*}

Therefore, the design of a low cost and robust cryptosystem for IoT is vital. While the main security streams have focused on the upper layers, the physical layer can also be leveraged to enhance security. In fact, reusing the physical layer features can decrease additional energy cost for security. As shown in Figure~\ref{fig:taxonomy}, security enhancement at the physical layer can be twofold. Firstly, information-theoretic security, also known as physical layer security (PLS), exploits the unpredictable features of wireless channels, such as fading; therefore, the system will not be compromised no matter how powerful the attackers are~\cite{zhou2013physical,he2013wireless,mukherjee2014princples,liu2017physical}.
PLS transmission techniques achieve security through artificial noise~\cite{goel2008guaranteeing}, jamming~\cite{ma2010approach}, or beamforming~\cite{mukherjee2011robust}, etc.
However, many PLS transmission schemes are not practical yet because they require complex coding and/or the perfect/imperfect channel state information (CSI) of the receiver and/or eavesdroppers~\cite{zou2016survey}. On the other hand, physical layer key generation, an active branch of PLS, is implementable because the legitimate users are able to agree on the same key from the noisy channel estimation~\cite{zhang2016review}, which can be used as an alternative to PKC in many circumstances. Secondly, moving the encryption to the physical layer can protect the entire physical layer packet and thus the wireless connection is secured from many passive and active attacks.

\begin{figure*}[!t]
	\centering
		\includegraphics[width=6in]{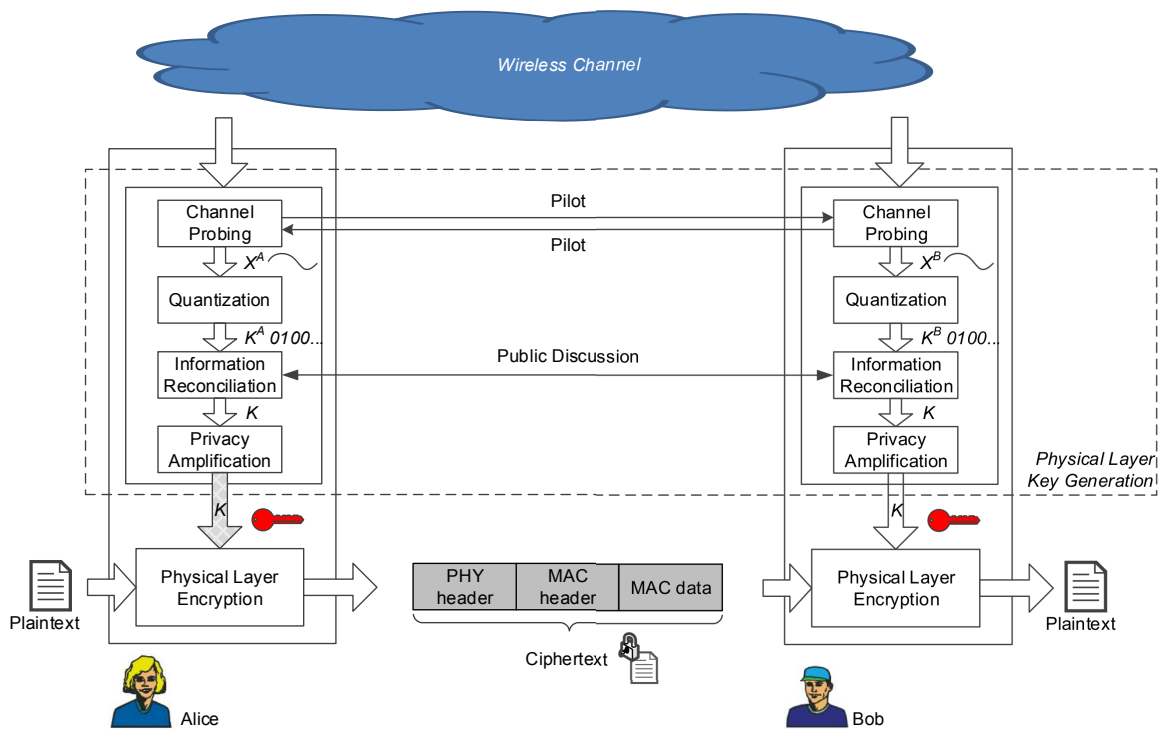}
	\caption{Key generation and PLE-based cryptosystem. The gray blocks represent the encrypted data.}
		\label{fig:KeyGen_PLE}
\end{figure*}

Recently, a new hybrid approach considers how we can deploy cryptosystems directly into the physical layer and integrates information-theoretical security and computational security schemes, which are constructed by physical layer key generation and physical layer encryption (PLE), as shown in Figure~\ref{fig:KeyGen_PLE}.
Alice and Bob carry out wireless transmission over the noisy channel using pilot signals. They are able to exploit common information of wireless channels and agree on the same cryptographic key through the key generation protocol consisted of channel probing, quantization, information reconciliation, and privacy amplification. The key is then fed to the PLE, which performs encryption operations at the modulation stages of the physical layer, and protects the IoT wireless transmission.
Their integration offers a good example of how information-theoretic security schemes and computational security schemes can work together to protect IoT systems.
Security countermeasures from the physical layer are lightweight and offer protection to the wireless transmission, and therefore are advantageous over conventional upper layer encryption-based security primitives.

There have been survey papers on the PLS transmission~\cite{mukherjee2015physical} and key generation~\cite{zeng2015physical,zhang2016review} to protect IoT. However, PLS transmission is limited in practical implementation and a survey on integration of key generation and encryption has never been reported. 
This article aims to provide an overview on the recent progress of this promising hybrid physical layer cryptosystem, with a focus on the practical implementation and algorithm prototyping.

The rest of this article is organized as follows. The wireless technologies used in IoT are introduced in Section~\ref{sec:wireless}. We then describe the physical layer key generation in Section~\ref{sec:key_gen} and PLE in Section~\ref{sec:PLE}.
Finally, we propose some future research directions in Section~\ref{sec:challenges} that make securing IoT from the physical layer more robust and pervasive. Section~\ref{sec:conclusion} concludes the article.

\section{Wireless Technologies for IoT and Their Security Countermeasures}\label{sec:wireless}
IoT aims to connect everything together and wireless communication is seen as the best option in order to avoid installation costs while enabling ubiquitous connection. IoT devices are normally tiny, embedded, and battery-powered, and thus communicate with each other through various low-power wireless communication technologies.
This section introduces several popular wireless technologies, including IEEE 802.15.4 (Zigbee), BLE, IEEE 802.11, and LoRaWAN.

IEEE 802.15.4 defines the physical and MAC layer protocols while Zigbee is based on IEEE 802.15.4 and includes high layer protocols. It runs at an unlicensed industrial, scientific and medical (ISM) 2.4~GHz frequency and uses direct sequence spread spectrum (DSSS) as the physical layer modulation. IEEE 802.15.4 is energy efficient and supports a data rate of up to 250 kbps, which is quite suitable for applications with limited data exchange requirements. It has been used extensively in wireless sensor networks (WSNs), especially in industrial applications.

BLE, also known as Bluetooth Smart, was standardized in 2010 as Bluetooth Core Specification Version 4.0. BLE runs at 2.4~GHz frequency and uses frequency hopping spread spectrum (FHSS) to combat frequency interference. It supports short range communications (50 to 100~m) and a data rate of 1~Mbps. In addition, BLE consumes extremely low energy and can run for months on standard coin-cell batteries. It is suited for IoT applications such as wearable devices and it is predicted by the Bluetooth special interest group  that more than 90\% smartphones will support BLE by 2018~\cite{bluetooth2015}.

IEEE 802.11 families are the most popular wireless local area network (WLAN) standards working at 2.4/5~GHz. They include IEEE 802.11a/b/g/n and are supported almost by all smartphones, laptops, tablets, etc.
IEEE 802.11 can be used in the smart home applications to provide large amounts of data transfer and as most houses are already covered by IEEE 802.11, installation costs can be avoided.
Whilst legacy IEEE 802.11 standards may not be suitable for many lightweight IoT applications, IEEE 802.11ah was recently announced by the Wi-Fi alliance which has been developed explicitly for IoT. It works at sub 1~GHz bands and can cover large range communications. In addition, 802.11ah adopts narrower bandwidth and implements energy efficient protocols to extend the sensors' battery life. It is also optimized to support large groups of stations or sensors that cooperate to share the signals.

LoRaWAN is an emerging low power wide area network (WAN) technology with the first specification released in June, 2015~\cite{lora}. It also runs at sub 1 GHz and employs chirp spread spectrum. It supports long range coverage ($>15$ km), millions of users, and low power consumption (up to ten years), and therefore is extremely suitable for low cost IoT devices.

IEEE 802.15.4, Bluetooth, and IEEE 802.11 systems usually handle the security at the data link/MAC layer. For example, an AES block cipher is used to protect the link layer of IEEE 802.15.4 and Bluetooth systems. In IEEE 802.11 systems, an MAC layer encryption scheme named WPA has been designed and implemented using AES. LoRa implements the security countermeasures by encryption at network and application layers.

A summary and comparison of the wireless technologies for the IoT is given in Table~\ref{tab:wireless}.
\begin{table*}[!t]
	\centering
		\caption{Wireless technologies for the IoT}
			\label{tab:wireless}
\scalebox{0.93}{
\begin{tabular}{|L{2cm}|l|l|l|L{3.8cm}|L{4.2cm}|}
    \hline
    Technique & Frequency & Range & Data Rate & Security Countermeasure & Applications \bigstrut\\
    \hline
    IEEE 802.15.4 (Zigbee) & 2.4 GHz & 10 to 100 m & 250 kps & AES in MAC layer & WSN, industrial, environment, and healthcare monitoring \bigstrut\\
    \hline
    BLE & 2.4 GHz & 50 to 150 m & 1 Mbps & AES in link layer & Wearable devices, smartphones \bigstrut\\
    \hline
    IEEE 802.11 a/b/g/n & 2.4 or 5 GHz & 50 m  & $>$ 100 Mbps & \multirow{2}[3]{3.8cm}{WPA in MAC layer (with AES implemented)} & Smart home, entertainment \bigstrut\\
\cline{1-4}\cline{6-6}    IEEE 802.11 ah & sub 1 GHz & 1 km  & 150 Kbps &       & Smart city, smart grid,  smart home, healthcare,  \bigstrut[t]\\\hline
    LoRaWAN  & sub 1 GHz & $>$ 15 km &  0.3 kbps to 50 kbps & Encryption at network and application layer & Machine-to-machine, smart city, and industrial applications \bigstrut[b]\\
    \hline
    \end{tabular}%
}		
  \label{tab:addlabel}%
\end{table*}%

\section{Physical Layer Key Generation}\label{sec:key_gen}
Key generation from the randomness of wireless channels has been receiving much research interest~\cite{zhang2016review}, as it is well-suited for establishing cryptographic keys as an alternative to PKC in IoT applications~\cite{zenger2014novel}.
As shown in Figure~\ref{fig:KeyGen_PLE}, firstly, key generation exploits unpredictable but characteristic features of the wireless channel, and is thus information-theoretically secure~\cite{ahlswede1993common,maurer1993secret}.
Secondly, it can be carried out between a pair of users with no aid from a third user, while a secured PKI is always required for PKC. Finally, key generation is lightweight and uses limited resources as all of the operations are not complicated and thus meets the low computation capacity of IoT devices.
Zenger~\textit{et al.} implemented their key generation scheme in a 32-bit ARM Cortex M3 processor (EFM32GG-STK3700) and an 8-bit Intel MCS-51 and showed the resource and energy consumption to be very low~\cite{zenger2016authenticated}. The authors also implemented a lightweight PKC, elliptic curves Diffie-Hellman key generation (ECDH), as a comparison. As shown in Table~\ref{tab:resource_energy}, taking the implementation in 32-bit ARM processor as an example, the ECDH requires 5.73 times more code, 128.26 times more cycles, and consumes 41.52 times more energy, than that of the key generation protocol, respectively. Therefore, key generation from wireless channels is extremely suitable for low cost IoT devices.

\begin{table*}[!t]
  \centering
  \caption{Resource and energy consumption comparison between key generation and ECDH}
\scalebox{0.85}{	
    \begin{tabular}{|l|r|r|r|r|r|r|r|}
    \hline
    \multirow{2}[4]{*}{Protocol} & \multicolumn{1}{c|}{\multirow{2}[4]{*}{Platform}} & \multicolumn{1}{l|}{\multirow{2}[4]{*}{Architecture}} & \multicolumn{2}{c|}{Resources} & \multicolumn{3}{c|}{Energy} \bigstrut\\
\cline{4-8}          & \multicolumn{1}{c|}{} & \multicolumn{1}{l|}{} & \multicolumn{1}{l|}{Code Size (kb)} & \multicolumn{1}{l|}{Cycles} & \multicolumn{1}{l|}{Computation (mJ)} & \multicolumn{1}{l|}{Communication (mJ)} & \multicolumn{1}{l|}{Total (mJ)} \bigstrut\\
    \hline
    Key generation & \multicolumn{1}{l|}{ARM Cortex-M3} & 32-bit & 1.033 & 302,297 & 2.246 & 0.187 & 2.433 \bigstrut\\
    \hline
    Key generation & \multicolumn{1}{l|}{Intel MCS-51} & 8-bit & 1.137 & 1,345,205 & 5.206 & 0.187 & 5.393 \bigstrut\\
    \hline
    ECDH  & \multicolumn{1}{l|}{ARM Cortex-M3} & 32-bit & 5.918 & 38,774,000 & 100.96 & 0.064 & 101.024 \bigstrut\\
    \hline
    ECDH  & \multicolumn{1}{l|}{Intel MCS-51} & 8-bit & 8.749 & 1,734,400,000 & 528.45 & 0.064 & 528.514 \bigstrut\\
    \hline
    \end{tabular}%
}		
  \label{tab:resource_energy}%
\end{table*}%

Key generation works well in a dynamic wireless communication system, and is built on three principles.
\begin{itemize}
	\item \textit{Channel reciprocity} means the channel responses of the forward and backward links are the same, which is the basis for key generation. When two users measure the same channel parameters at the same frequency in a time-division duplex (TDD) mode, the measurements at Alice and Bob are impacted by the non-simultaneous sampling and noise. However, a high correlation between channel measurements of Alice and Bob can still be maintained and eligible for key generation in a slow fading channel, as demonstrated in many practical experiments~\cite{mathur2008radio,jana2009effectiveness,zhang2016spawc,zhang2016experimental}.
  \item \textit{Temporal variation} indicates that there is randomness residing in the dynamic channel\footnote{In the urban area, the interference may be chaotic, because of the densely deployed access points~\cite{kajita2016wi}. The interference will impact the channel measurements accuracy but will not affect randomness nature of the wireless link between users. In addition, the statistical features of the channel may be deterministic~\cite{chin2014wireless,santana2016adaptive}, but key generation is exploiting the instantaneous channel variation, which is random in nature.}, which ensures the extracted keys are random. A random key will make the cryptographic applications robust against  attacks such as brute force. 
	\item \textit{Spatial decorrelation} implies that when located a half-wavelength away from the legitimate users, the eavesdropper experiences an uncorrelated channel compared to that between Alice or Bob, guaranteeing the security of the key generation. When the system works at 2.4~GHz, a half-wavelength is about 6~cm, which is quite short.
\end{itemize}
These principles have been theoretically modeled and analyzed in~\cite{zhang2016efficient,zhang2017on} and experimentally validated in~\cite{zhang2016experimental,zhang2016spawc}.

\subsection{Procedure}
Key generation involves channel probing, quantization, information reconciliation, and privacy amplification, as shown in Figure~\ref{fig:KeyGen_PLE}. Without loss of generality, Alice is selected as the initiator of the key generation process.

In the channel probing step, the randomness residing in the temporal~\cite{mathur2008radio,jana2009effectiveness,zhang2016efficient}, frequency~\cite{liu2013fast,xi2014KEEP,zhang2016efficient,peng2017secret}, and spatial~\cite{zeng2010exploiting,wallace2010automatic,chen2011secret,jorswieck2013secret} domains can be extracted by measuring the
channel parameters such as the received signal strength (RSS) and CSI, etc.
In particular, at time $t_A$ Alice sends a public pilot signal to Bob who will measure the channel parameter as $X^B$. Then, at time $t_B = t_A+\tau$, Bob also sends a public pilot signal to Alice who will measure the same channel parameter and store it as $X^A$. Alice and Bob will repeat the above channel sampling until they get enough measurements to generate a full set of keys\footnote{The key length is determined by the cryptographic applications. For example, the key length of AES can be 128-bit, 192-bit, or 256-bit.}.
It is worth noting that in this step, users adopt a public pilot signal to measure the channel but do not try to exchange message secretly. It is possible that some of the probe packets are not successful because of the poor channel condition, which results in a mismatch between the pairing of the measurements of Alice and Bob. This can be solved by exchanging and comparing the timestamps of the measurements, and keeping the records with the common timestamps. In TDD mode, the common timestamp does not necessarily indicate the timestamps with the exact same value, but their difference should be the sampling delay $\tau$ . For example, Alice will send her recorded timstamps to Bob, who will compare his timestamps and keep the common ones. Bob will then send his censored timestamps to Alice and she will also only keep the common ones, which will finally enable Alice and Bob to have the paired measurements.
The exchange does not reveal any useful information to eavesdroppers.

In the second step, both Alice and Bob will convert the analog measurements into binary sequences using quantization schemes. Mean and standard deviation-based quantizer~\cite{mathur2008radio} (Algorithm~\ref{alg:mean}) and cumulative distribution functions (CDF)-based quantizer~\cite{patwari2010high} (Algorithm~\ref{alg:cdf}) are two popular quantizers. In Algorithm~\ref{alg:mean}, $\mu_{X^{u}}$ is the mean value of $X^u$, $\sigma_{X^{u}}$ is the standard deviation of $X^u$, $\alpha$ is used to adjust the threshold, and $n$ is the number of the channel measurements. The design of quantizer relies on the selection of threshold and quantization level (QL). CDF-based quantization is able to obtain the same proportion of 0s and 1s as it can adaptively adjust the threshold, which is at the cost of increased complexity. The computational complexity of calculating the mean and variance is $\mathcal{O}(n)$. When calculating CDF, one key step is sorting the measurements, whose complexity is $\mathcal{O}\big(n\log(n)\big)$, which requires more computation than the calculation of the mean and variance. 
A performance comparison of quantization schemes is reported in~\cite{zenger2015security}.

\begin{algorithm}[!t]
\caption{Mean and standard deviation-based quantization}
\begin{algorithmic}[1]
\INPUT $X^u$ \Comment{Channel measurement, RSS or CSI}
\OUTPUT $K^u$ \Comment{Key}
\State{$\eta_+^u =  \mu_{X^{u}} + \alpha \times \sigma_{X^{u}}$} 
\Comment{$\eta_+^u$ is the positive threshold.}
\State{$\eta_-^u =  \mu_{X^{u}} - \alpha \times \sigma_{X^{u}}$} 
\Comment{$\eta_-^u$ is the negative threshold.}
\For{$i \gets 1 \:\:\:\: \textbf{to} \:\:\:\:  n$}
\If{$X^{u}(i)>\eta_+^u$}
    \State{$K^{u}(i) = 1$}
\ElsIf{$X^{u}(i) < \eta_-^u$}
    \State{$K^{u}(i) = 0$}
\Else		
\State{$X^{u}(i)$ dropped}
\EndIf
\EndFor
\end{algorithmic}
\label{alg:mean}
\end{algorithm}

\begin{algorithm}[!t]
\caption{CDF-based quantization}
\begin{algorithmic}[1]
\INPUT $X^u$ \Comment{Channel measurement, RSS or CSI}
\INPUT \textit{QL} \Comment{Quantization level}
\OUTPUT $K^u$ \Comment{Key}
\State{$ F(x) = \text{Pr}(X^{u}<x)$} \Comment{CDF calculation}

\State{$\eta_0^u = -\infty$} \Comment{Threshold}
\For{$j \gets 1\:\:\:\: \textbf{to} \:\:\:\: 2^\textit{QL}-1$}
\State{$\eta_j^u = F^{-1}(\frac{j}{2^\textit{QL}})$}\Comment{Threshold}
\EndFor
\State{$\eta_{2^\textit{QL}}^u = \infty$}

\State{Construct Gray code $b_j$ and assign them to different intervals $[\eta_{j-1}^u, \eta_{j}^u]$}

\For{$i \gets 1\:\:\:\: \textbf{to} \:\:\:\: n$}
\If{$\eta_{j-1}^u \leq X^{u}(i) < \eta_j^u $}
\State{$K^{u}(i,\textit{QL}) = b_j$ }

\EndIf

\EndFor
\end{algorithmic}
\label{alg:cdf}
\end{algorithm}

In practical measurements, due to the half-duplex nature of the most commercial hardware platforms and the independent hardware noise, channel measurements of Alice and Bob, i.e., $X^A$ and $X^B$, will not be identical, thus resulting in a disagreement between $K^A$ and $K^B$. In the information reconciliation stage, Alice and Bob will leverage the error correction code (ECC) to reach an agreement, which is achieved via public discussion by exchanging information such as the syndrome. Secure sketch~\cite{dodis2008fuzzy} is a popular key reconciliation technique and is given as an example in Algorithm~\ref{alg:ss}. A comprehensive survey on information reconciliation techniques can be found in~\cite{huth2016information}. Finally, privacy amplification is employed to eliminate the information revealed to eavesdroppers, which can be implemented using hash functions~\cite{zhang2016review}.

\begin{algorithm}[!t]
\caption{Secure sketch}
\begin{algorithmic}[1]
\INPUT $K^A$, $K^B$ \Comment{Quantized keys of Alice and Bob}
\INPUT $C$ \Comment{ECC set shared by Alice and Bob}
\OUTPUT $K^A$, $K^{B'}$ \Comment{Reconciled key}
\State{Alice randomly selects $c$ from an ECC set $C$ }
\State{Alice calculates $s = \text{XOR}(K^A,c)$}
\State{Alice transmits $s$ to Bob through a public channel}
\State{Bob receives $s$}
\State{Bob calculates $c^B = \text{XOR}(K^B,s)$}
\State{Bob decodes $c^B$} to get $c$ 
\State{Bob calculates $K^{B'} = \text{XOR}(c,s) = K^A$}~\Comment{Alice and Bob agree on the same key}
\end{algorithmic}
\label{alg:ss}
\end{algorithm}

\subsection{Application}
Due to its lightweight  feature, this form of key generation has strong potential to provide the security for IoT. It has been applied in many wireless technologies, such as IEEE 802.11, IEEE 802.15.4, Bluetooth, etc., with many prototypes/implementations reported, see~\cite{zhang2016review}.

IEEE 802.11 is the most popular technique for the key generation implementation as the technique is widely adopted in our daily life. The work in~\cite{mathur2008radio} is one of the first and important papers that implemented key generation protocol. The authors generated keys from the peak of channel impulse response (CIR) using an 802.11 compatible field-programmable gate array (FPGA)-based platform, and also from RSS with a commercial Wi-Fi network interface card (NIC). However, the key generation rate is rather limited, i.e., about 1~bps, since the authors only extracted keys from coarse-grained channel parameter. Orthogonal frequency-division multiplexing (OFDM) is employed by IEEE 802.11a/g/n/ah, which can provide fine-grained CSI in both time and frequency domain and significantly improve the key generation performance~\cite{zhang2016efficient,liu2013fast}.

A key generation system using wearable devices with IEEE 802.15.4 is implemented in~\cite{Ali2014Eliminating}.
Channel measurements are carried out along with data transmission, in other words, no dedicated transmission is incurred for key generation.
This avoids the additional energy burden required by key generation, which can significantly save power consumption as the radio transmission is always the dominant~\cite{trappe2015low}.
In addition, a low cost filter is employed to improve the signal cross-correlation, which helps the system reach an agreement as high as 99.8\%~\cite{Ali2014Eliminating}.
Since there is not much data transmission required by wearable devices, the system takes about half an hour to generate 128-bit keys. The duration is acceptable as it still meets the requirement. For example, Wi-Fi recommends to refresh the session key every hour.

Key generation has also been applied in Bluetooth systems~\cite{premnath2014secret}. The authors implemented their system in two Google Nexus One smartphones and sampled RSS with experiments in indoor and outdoor environments.
Random frequency hopping was employed to combat the interference from other wireless networks running at the ISM bands. It has also been demonstrated by experiments that Bluetooth-based key generation can be carried out using much lower transmit power (3~dBm) with a performance comparable to that of Wi-Fi-based system, which is desirable for IoT devices.

\section{Physical Layer Encryption}\label{sec:PLE}
Modern communication systems employ a layered protocol stack to organize communication functions and most of the current security methodologies are applied at the MAC layer and above. The physical layer is the lowest layer of the protocol stack and was designed originally to modulate data for transmission but without any security considerations. This section introduces some recent ongoing encryption schemes implemented at the physical layer, which protects the entire physical layer packet. PLE schemes are lightweight as they do not introduce additional complexity, therefore are quite suitable for IoT applications.

\subsection{Procedure}
\begin{figure*}[!t]
	\centering
		\includegraphics[width=7in]{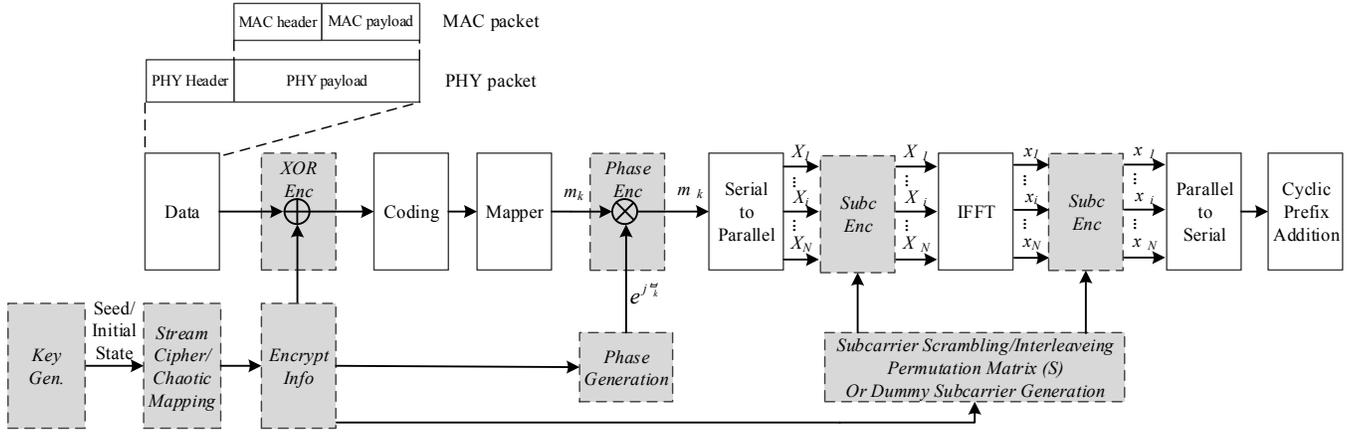}
	\caption{Physical layer encryption schemes in OFDM. Gray modules are added for encryption.}
		\label{fig:PLE}
\end{figure*}

The data payload undergoes several physical layer modulation stages, such as channel coding, mapping, inverse fast Fourier transform (IFFT) operation (for OFDM systems), etc. PLE can be applied by encrypting the data flow in these physical layer modulation stages. Some PLE schemes applicable for OFDM systems are shown in Figure~\ref{fig:PLE}, including XOR encryption~\cite{huo2015xor}, phase encryption~\cite{reilly2009noise,ma2010secure,huo2015xor}, and OFDM subcarriers encryption~\cite{khan2007secure,tseng2007ofdm,zhang2011secure,li2013secure,li2014dynamic,li2015eavesdropping,zhang2016design}. The user first generates the encryption information using the output of stream cipher or chaotic mapping. Based on the adopted encryption scheme, the encryption information is used to calculate phase rotation, dummy subcarrier locations, or subcarrier scrambling/interleaving permutation, etc., which is then used to protect the corresponding modulation stage. The detailed calculation step will shown in Section~\ref{sec:prototype}.  The seed for the stream cipher or the initial state of the chaotic map can be shared between legitimate users using the key generation discussed in the last section.

The entire packet is protected. The encryption of the physical layer payload, i.e., the MAC layer packet, will secure the MAC layer content, including the MAC header. In addition, the protection of the physical layer header can prevent eavesdroppers from carrying out functions of synchronization and channel estimation, significantly increasing the processing overheads for the eavesdropper~\cite{zhang2016design}.

\subsection{Algorithm Prototype}\label{sec:prototype}
The PLE design is determined by the wireless technologies which employ different physical layer modulations. In this section, we introduce several PLE prototypes based on the modulation stage that they have encrypted.

XOR encryption is the most straightforward and lightweight scheme and can be implemented in hardware in a very efficient manner. As XOR is a bitwise operation, it usually happens before coding, as shown in Figure~\ref{fig:PLE}. This scheme is applicable to all the wireless technologies as the data passed from the MAC layer is always in binary form. However, it is implemented at the beginning of modulation stages and does not randomize the physical layer waveform, which results in a weaker protection~\cite{zhang2016design}.

Phase encryption can also be applied  as long as phase-shift keying or quadrature amplitude modulation is used~\cite{reilly2009noise,ma2010secure,huo2015xor}. As shown in Figure~\ref{fig:PLE}, phase encryption occurs after symbol mapping and the constellation symbols are not in binary values any more.
The encrypted constellation symbols $m_k'$ can be denoted as
\begin{align}
	m_k' = m_k e^{j\theta_k} + n_k,
\end{align}
where $m_k$ is the constellation symbols, $\theta_k$ is the rotation angle, and $n_k$ is the random noise. $\theta_k$ is generated according to the key sequence and then used to rotate the constellation symbols. In order to create a denser encrypted constellation, more key bits are required to generate rotation angles, which increases the key-to-data ratio\footnote{Key-to-data ratio is defined as the number of key bits needed to encrypt one bit plaintext, which is a key metric of PLE.}. Random noise, $n_k$,  can be deliberately added to the rotated symbols to make it even more difficult for the eavesdroppers to demodulate the ciphertext~\cite{reilly2009noise,ma2010secure}. The implementation of this technique is also  efficient because the main resource is a multiplier and related control circuits.

The OFDM technique modulates data onto multiple orthogonal subcarriers/frequencies and can significantly increase the data rate, providing an additional domain to protect the data. The parallel input data $X$ can be scrambled in frequency domain before IFFT operation~\cite{khan2007secure,tseng2007ofdm,zhang2011secure}, which can be given as
\begin{align}
	X' = X S_f,
\end{align}
where $S_f$ is the frequency scrambling matrix, or the IFFT output data can be scrambled in the time domain~\cite{li2013secure} and written as
\begin{align}
	x' = x S_t,
\end{align}
where $S_t$ is the time scrambling matrix.
Scramble-based schemes can bring a large search space. However, it may result in a high computational complexity as matrix operations are required, which may not be suitable for low cost devices~\cite{zhang2016design}.

Different from above OFDM schemes that scramble all the data subcarriers, the work in~\cite{li2014dynamic,li2015eavesdropping} interleaves only part of the subcarriers. In particular, the scheme in~\cite{li2014dynamic} selects a subset of the subcarriers whose phase is larger than the threshold, and then interleaves their real and imaginary components of the symbols. The method in~\cite{li2015eavesdropping} selects a subcarrier subset based on the CSI, and then interleaves these subcarriers according to the descending order of their channel amplitudes. Encryption usually involves mathematical operations, e.g., XOR operation, between the plaintext and key sequence, but here the concept applies more generally to the data manipulation according to the common secret information. In addition, the authors use channel information as encryption information directly without resorting to stream ciphers, which requires a careful design of the interleaving pattern because of the channel estimation errors at transmitters and receivers.

While the standard OFDM systems use all the data subcarriers for data transmission, some subcarriers can also be reserved to transmit dummy data, i.e., rubbish information, for obfuscation~\cite{zhang2016design}. Due to the introduction of dummy subcarriers, there is a trade-off between the security and data rate, but it has been demonstrated in~\cite{zhang2016design} that it is worthwhile as there are many subcarriers and the data rate is usually only slightly reduced. In addition, the preambles are encrypted in~\cite{zhang2016design} so the entire packet is protected.

The above schemes protect different physical layer modulation stages, which lead to distinctions on the security level, complexity, etc. For example, XOR and phase encryption are easier to implement but provide less strong protection. On the other hand, scrambling-based schemes may require matrix operations, including matrix multiplication and inversion, which result in a higher computation complexity.
A detailed comparison in terms of search space to the brute force attack, key rate, and complexity of the above schemes can be found in~\cite{zhang2016design}.

\subsection{Practical Implementation}
To the best of the authors' knowledge, there has been only one paper which has implemented a physical layer phase encryption IEEE 802.15.4 transceiver and RC4 to generate the key sequence~\cite{nain2017secure}. The work in~\cite{nain2017secure}  first validated the design using FPGA technology and then implemented the system in application-specific integrated circuit (ASIC) using UMC 0.18 $\mu m$ complementary metal-oxide-semiconductor (CMOS) technology. The security enhancement, including the RC4 and phase encryption/decryption, results in a 26\% increase on the gate counts compared to a standard 802.15.4 transceiver, which is a reasonable overheard for security.

\section{Future Work Suggestions}\label{sec:challenges}
Although there have been prototypes/demonstrations of the above physical layer-based security countermeasures, research is still needed to make these schemes more robust and pervasive. In this section, we suggest some future research directions in securing the IoT from the physical layer.

\subsection{Physical Layer Key Generation}
Most current commercial platforms work in half-duplex mode, and the keying nodes have to measure the channel alternately in different time instances. Key generation in this setting is only applicable to slow fading channels in order to get a highly correlated measurements between users. Therefore, key generation in fast fading channels is very challenging, which limits its application, e.g. in vehicular communications. Work in~\cite{wan2016exploiting} and~\cite{zhu2017using} designed key generation systems with the maximum vehicle speed tested as 20~mph and 50~mph, respectively, but their key generation rates are limited, e.g., 5 bit/s in~\cite{zhu2017using}. In addition, work in~\cite{li2017efficient} tested their algorithms in indoor environment only. There is also some simulation work, e.g.,~\cite{abdelgader2014secret,abdelgader2017exploiting}. Their performance in the practical fast fading channels remains unknown.
This topic is thus still require more efforts, e.g., by using full-duplex hardware~\cite{vogt2016practical}. 

Efficient group and pairwise key generation are essential to assist secure broadcast and unicast transmission in a large scale IoT network. A fusion center broadcasts signals to the network users, which requires a pre-establishment of a common session key. The devices may also exchange unicast packets between each other, and private keys between pairs of users are required. In ad hoc IoT, many users may join and leave the network frequently, therefore robust and efficient schemes to update the session key and private keys are required.
There have been several group key generation protocols reported, e.g., a time-slotted round-trip phase-based scheme~\cite{wang2011fast}, RSS-based protocols for star and chain topologies~\cite{liu2014group}, and group key generation for mesh topology~\cite{thai2015secret}. However, the scalability (with the size of the group) and efficiency of the above protocols are limited and more research effort is required.

Although key generation is able to achieve information-theoretic security, in practice the security performance requires special attention. For example, when there is a strong line-of-sight, the spatial decorrelation may not hold any more, which makes the system vulnerable to passive eavesdropping~\cite{zhang2016experimental,zenger2016passive}. Key generation is also subject to active attacks~\cite{zafer2012limitations,jin2015physical}, which will result in less efficient or even unsuccessful key generation. It is thereof very important to design key generation techniques secure from passive eavesdropping and robust to active jamming. In addition, the majority of the research focuses on the indoor and/or mobile channels, while in an outdoor or static environment, the channel randomness is rather limited. A less random key will expose the cryptographic systems to brute force attack and should be always avoided.

\subsection{Physical Layer Encryption}
PLE applies encryption at the physical layer, and entails additional operations and hardware resources. No hardware implementation for PLE schemes has been reported except for those in~\cite{nain2017secure}.
The additional operations will introduce latency in the critical path and may not meet the timing requirements of the current MAC protocol. Therefore, a cross-layer design between the physical and MAC layer is necessary.

The keys generated are usually fed to a stream cipher to produce pseudo random numbers to encrypt plaintext.
In scenarios where keys can be generated fast or only very small amount of data exchange is required, the keys generated can be used to encrypt the data directly, rather than be used as the seed for stream cipher. Key generation and PLE is then integrated as a one-time pad scheme to offer perfect Shannon secrecy, which can provide strongest protection ever. However, the practical security performance and implementation requires further investigation.

\section{Conclusion}\label{sec:conclusion}
This article has provided an overview on securing wireless communications of IoT applications from the physical layer. We have introduced two security techniques, namely, physical layer key generation and physical layer encryption. For each, we have discussed their features and applications by a special consideration of IoT devices' low power and low cost features.
The remaining challenges of how to make these schemes more robust and pervasive have also been proposed.
Unlike previous work, this article has focused on practical prototypes/implementations, thus offering insights for their applications in the IoT to enhance wireless security.


\begin{thebibliography}{10}
\providecommand{\url}[1]{#1}
\csname url@samestyle\endcsname
\providecommand{\newblock}{\relax}
\providecommand{\bibinfo}[2]{#2}
\providecommand{\BIBentrySTDinterwordspacing}{\spaceskip=0pt\relax}
\providecommand{\BIBentryALTinterwordstretchfactor}{4}
\providecommand{\BIBentryALTinterwordspacing}{\spaceskip=\fontdimen2\font plus
\BIBentryALTinterwordstretchfactor\fontdimen3\font minus
  \fontdimen4\font\relax}
\providecommand{\BIBforeignlanguage}[2]{{%
\expandafter\ifx\csname l@#1\endcsname\relax
\typeout{** WARNING: IEEEtran.bst: No hyphenation pattern has been}%
\typeout{** loaded for the language `#1'. Using the pattern for}%
\typeout{** the default language instead.}%
\else
\language=\csname l@#1\endcsname
\fi
#2}}
\providecommand{\BIBdecl}{\relax}
\BIBdecl

\bibitem{atzori2010internet}
L.~Atzori, A.~Iera, and G.~Morabito, ``The {Internet of Things}: A survey,''
  \emph{Computer Networks}, vol.~54, no.~15, pp. 2787--2805, 2010.

\bibitem{al2015internet}
A.~Al-Fuqaha, M.~Guizani, M.~Mohammadi, M.~Aledhari, and M.~Ayyash, ``{Internet
  of Things: A survey on enabling technologies, protocols, and applications},''
  \emph{{IEEE} Commun. Surveys Tuts.}, vol.~17, no.~4, pp. 2347--2376, Fourth
  Quarter 2015.

\bibitem{stankovic2014research}
J.~A. Stankovic, ``Research directions for the internet of things,''
  \emph{{IEEE} Internet Things J.}, vol.~1, no.~1, pp. 3--9, 2014.

\bibitem{walport2014iot}
M.~Walport, ``The {Internet of Things}: Making the most of the second digital
  revolution, {A} report by the {UK} government chief scientific adviser,''
  Tech. Rep., December 2014,
  \url{https://www.gov.uk/government/uploads/system/uploads/attachment_data/file/409774/14-1230-internet-of-things-review.pdf},
  Accessed on 22 June 2017.

\bibitem{mckinsey2015iot}
``The {Internet of Things}: Five critical questions,'' McKinsey Global
  Institute, August 2015,
  \url{http://www.mckinsey.com/industries/high-tech/our-insights/the-internet-of-things-five-critical-questions},
  Accessed on 22 June 2017.

\bibitem{nsf}
``A partnership to secure and protect the emerging {Internet of Things},''
  National Science Foundation, August 2015,
  \url{http://nsf.gov/news/news_summ.jsp?cntn_id=136104&org=NSF}, Accessed on
  22 June 2017.

\bibitem{ierc}
European Research Cluster on the Internet of Things,
  \url{http://www.internet-of-things-research.eu/}, Accessed on 22 June 2017.

\bibitem{epsrc2016iothub}
``New {Internet of Things} research hub announced,'' The Engineering and
  Physical Sciences Research Council, January 2016,
  \url{https://www.epsrc.ac.uk/newsevents/news/iotresearchhub/} , Accessed on
  22 June 2017.

\bibitem{nordrum2016iot}
A.~Nordrum, ``The {Internet} of fewer things,'' September 2016,
  \url{http://spectrum.ieee.org/telecom/internet/the-internet-of-fewer-things},
  Accessed on 22 June 2017.

\bibitem{cisco2011iot}
D.~Evans, ``Internet of things research study,'' Cisco, Tech. Rep., April 2011,
  \url{http://www.cisco.com/c/dam/en_us/about/ac79/docs/innov/IoT_IBSG_0411FINAL.pdf},
  Accessed on 22 June 2017.

\bibitem{grau2016iot}
A.~Grau, ``How to build a safer {Internet of Things},'' February 2015,
  \url{http://spectrum.ieee.org/telecom/security/how-to-build-a-safer-internet-of-things},
  Accessed on 22 June 2017.

\bibitem{hp2015iot}
``Internet of things research study,'' HP, Tech. Rep., November 2015, \url
  {https://www.hpe.com/h20195/v2/GetPDF.aspx/4AA5-4759ENN.pdf}, Accessed on 22
  June 2017.

\bibitem{zou2016survey}
Y.~Zou, J.~Zhu, X.~Wang, and L.~Hanzo, ``A survey on wireless security:
  Technical challenges, recent advances, and future trends,'' \emph{Proc.
  {IEEE}}, vol. 104, no.~9, pp. 1727--1765, September 2016.

\bibitem{granjal2015security}
J.~Granjal, E.~Monteiro, and J.~Sa~Silva, ``{Security for the Internet of
  Things: A survey of existing protocols and open research issues},''
  \emph{{IEEE} Commun. Surveys Tuts.}, vol.~17, no.~3, pp. 1294--1312, Third
  Quarter 2015.

\bibitem{rfc5246tls}
\BIBentryALTinterwordspacing
T.~Dieks and E.~Rescorla, ``{{The Transport layer security (TLS) protocol}},''
  Internet Requests for Comments, {RFC Editor}, {RFC} 5246, August 2008.
  [Online]. Available: \url{http://www.rfc-editor.org/rfc/rfc5246.txt}
\BIBentrySTDinterwordspacing

\bibitem{ieee2004security}
``Wireless lan medium access control (mac) and physical layer (phy)
  specifications: Amendment 6: Medium access control (mac) security
  enhancements,'' IEEE, Tech. Rep. 802.11i, July 2004.

\bibitem{trappe2015low}
W.~Trappe, R.~Howard, and R.~S. Moore, ``Low-energy security: Limits and
  opportunities in the {Internet of Things},'' \emph{{IEEE} Security Privacy},
  vol.~13, no.~1, pp. 14--21, January/February 2015.

\bibitem{cheng2017securing}
C.~Cheng, R.~Lu, A.~Petzoldt, and T.~Takagi, ``Securing the {Internet of
  Things} in a quantum world,'' \emph{{IEEE} Commun. Mag.}, vol.~55, no.~2, pp.
  116--120, February 2017.

\bibitem{rahbari2015secrecy}
H.~Rahbari and M.~Krunz, ``Secrecy beyond encryption: obfuscating transmission
  signatures in wireless communications,'' \emph{{IEEE} Commun. Mag.}, vol.~53,
  no.~12, pp. 54--60, December 2015.

\bibitem{zhou2013physical}
X.~Zhou, L.~Song, and Y.~Zhang, Eds., \emph{Physical layer security in wireless
  communications}.\hskip 1em plus 0.5em minus 0.4em\relax CRC Press, 2013.

\bibitem{he2013wireless}
B.~He, X.~Zhou, and T.~D. Abhayapala, ``Wireless physical layer security with
  imperfect channel state information: A survey,'' \emph{ZTE Communications},
  vol.~11, no.~3, p. 11–19, September 2013.

\bibitem{mukherjee2014princples}
A.~Mukherjee, S.~Fakoorian, J.~Huang, and A.~Swindlehurst, ``Principles of
  physical layer security in multiuser wireless networks: A survey,''
  \emph{{IEEE} Commun. Surveys Tuts.}, vol.~16, no.~3, pp. 1550--1573, Third
  Quarter 2014.

\bibitem{liu2017physical}
Y.~Liu, H.-H. Chen, and L.~Wang, ``Physical layer security for next generation
  wireless networks: Theories, technologies, and challenges,'' \emph{{IEEE}
  Commun. Surveys Tuts.}, vol.~19, no.~1, pp. 347 -- 376, 2017.

\bibitem{goel2008guaranteeing}
S.~Goel and R.~Negi, ``Guaranteeing secrecy using artificial noise,''
  \emph{IEEE Trans. Wireless Commun.}, vol.~7, no.~6, 2008.

\bibitem{ma2010approach}
S.~Ma, M.~Hempel, Y.~L. Yang, and H.~Sharif, ``An approach to secure wireless
  communications using randomized eigenvector-based jamming signals,'' in
  \emph{Proc. 6th Int. Wireless Communications and Mobile Computing Conf.},
  Caen, France, July 2010, pp. 1172--1176.

\bibitem{mukherjee2011robust}
A.~Mukherjee and A.~L. Swindlehurst, ``Robust beamforming for security in mimo
  wiretap channels with imperfect csi,'' \emph{IEEE Trans. Signal Processing},
  vol.~59, no.~1, pp. 351--361, 2011.

\bibitem{zhang2016review}
J.~Zhang, T.~Q. Duong, A.~Marshall, and R.~Woods, ``Key generation from
  wireless channels: A review,'' \emph{IEEE Access}, vol.~4, pp. 614--626,
  March 2016.

\bibitem{mukherjee2015physical}
A.~Mukherjee, ``Physical-layer security in the {Internet of Things}: Sensing
  and communication confidentiality under resource constraints,'' \emph{Proc.
  {IEEE}}, vol. 103, no.~10, pp. 1747--1761, October 2015.

\bibitem{zeng2015physical}
K.~Zeng, ``Physical layer key generation in wireless networks: challenges and
  opportunities,'' \emph{IEEE Communications Magazine}, vol.~53, no.~6, pp.
  33--39, 2015.

\bibitem{bluetooth2015}
S.~Janiak, ``Three ways {Bluetooth}\textsuperscript{\textregistered} smart
  technology enables innovation for the {Internet of Things},'' January 2015,
  \url{http://blog.bluetooth.com/three-ways-bluetooth-smart-technology-enables-innovation-for-the-internet-of-things/},
  Accessed on 22 June 2017.

\bibitem{lora}
``{LoRa Alliance},'' \url{https://www.lora-alliance.org/}, Accessed on 22 June
  2017.

\bibitem{zenger2014novel}
C.~T. Zenger, M.-J. Chur, J.-F. Posielek, C.~Paar, and G.~Wunder, ``A novel key
  generating architecture for wireless low-resource devices,'' in \emph{Proc.
  Int. Workshop Secure Internet of Things (SIoT)}, Wroclaw, Poland, September
  2014, pp. 26--34.

\bibitem{ahlswede1993common}
R.~Ahlswede and I.~Csiszar, ``Common randomness in information theory and
  cryptography -- {Part I}: secret sharing,'' \emph{{IEEE} Trans. Inf. Theory},
  vol.~39, no.~4, pp. 1121--1132, 1993.

\bibitem{maurer1993secret}
U.~M. Maurer, ``Secret key agreement by public discussion from common
  information,'' \emph{{IEEE} Trans. Inf. Theory}, vol.~39, no.~3, pp.
  733--742, 1993.

\bibitem{zenger2016authenticated}
C.~T. Zenger, M.~Pietersz, J.~Zimmer, J.-F. Posielek, T.~Lenze, and C.~Paar,
  ``Authenticated key establishment for low-resource devices exploiting
  correlated random channels,'' \emph{Computer Networks}, vol. 109, pp.
  105--123, 2016.

\bibitem{mathur2008radio}
S.~Mathur, W.~Trappe, N.~Mandayam, C.~Ye, and A.~Reznik, ``Radio-telepathy:
  Extracting a secret key from an unauthenticated wireless channel,'' in
  \emph{Proc. 14th Annu. Int. Conf. Mobile Computing Networking (MobiCom)}, San
  Francisco, California, USA, September 2008, pp. 128--139.

\bibitem{jana2009effectiveness}
S.~Jana, S.~N. Premnath, M.~Clark, S.~K. Kasera, N.~Patwari, and S.~V.
  Krishnamurthy, ``On the effectiveness of secret key extraction from wireless
  signal strength in real environments,'' in \emph{Proc. 15th Annu. Int. Conf.
  Mobile Computing and Networking (MobiCom)}, Beijing, China, September 2009,
  pp. 321--332.

\bibitem{zhang2016spawc}
J.~Zhang, R.~Woods, T.~Q. Duong, A.~Marshall, and Y.~Ding, ``Experimental study
  on channel reciprocity in wireless key generation,'' in \emph{Proc. 17th IEEE
  Int. Workshop Signal Process. Advances in Wireless Commun. (SPAWC)},
  Edinburgh, UK, July 2016, pp. 1--5.

\bibitem{zhang2016experimental}
J.~Zhang, R.~Woods, T.~Q. Duong, A.~Marshall, Y.~Ding, Y.~Huang, and Q.~Xu,
  ``Experimental study on key generation for physical layer security in
  wireless communications,'' \emph{IEEE Access}, vol.~4, pp. 4464--4477,
  September 2016.

\bibitem{kajita2016wi}
S.~Kajita, T.~Amano, H.~Yamaguchi, T.~Higashino, and M.~Takai, ``Wi-fi channel
  selection based on urban interference measurement,'' in \emph{Proc. 13th Int.
  Conf. on Mobile and Ubiquitous Systems: Computing, Networking and Services},
  Hiroshima, Japan, November/December 2016, pp. 143--150.

\bibitem{chin2014wireless}
E.~Chin, D.~Chieng, V.~Teh, M.~Natkaniec, K.~Loziak, and J.~Gozdecki,
  ``Wireless link prediction and triggering using modified ornstein--uhlenbeck
  jump diffusion process,'' \emph{Wireless Networks}, vol.~20, no.~3, pp.
  379--396, 2014.

\bibitem{santana2016adaptive}
J.~A. Santana, E.~Mac{\'\i}as, {\'A}.~Su{\'a}rez, D.~Marrero, and V.~Mena,
  ``Adaptive estimation of wifi rssi and its impact over advanced wireless
  services,'' \emph{Mobile Networks and Applications}, pp. 1--13, 2016.

\bibitem{zhang2016efficient}
J.~Zhang, A.~Marshall, R.~Woods, and T.~Q. Duong, ``Efficient key generation by
  exploiting randomness from channel responses of individual {OFDM}
  subcarriers,'' \emph{{IEEE} Trans. Commun.}, vol.~64, no.~6, pp. 2578--2588,
  June 2016.

\bibitem{zhang2017on}
J.~Zhang, B.~He, T.~Q. Duong, and R.~Woods, ``On the key generation from
  correlated wireless channels,'' \emph{{IEEE} Commun. Lett.}, vol.~21, no.~4,
  pp. 961--964, 2017.

\bibitem{liu2013fast}
H.~Liu, Y.~Wang, J.~Yang, and Y.~Chen, ``Fast and practical secret key
  extraction by exploiting channel response,'' in \emph{Proc. 32nd IEEE Int.
  Conf. Comput. Commun. (INFOCOM)}, Turin, Italy, April 2013, pp. 3048--3056.

\bibitem{xi2014KEEP}
W.~Xi, X.~Li, C.~Qian, J.~Han, S.~Tang, J.~Zhao, and K.~Zhao, ``{KEEP: Fast
  secret key extraction protocol for D2D communication},'' in \emph{Proc. 22nd
  IEEE Int. Symp. of Quality of Service (IWQoS)}, Hong Kong, May 2014, pp.
  350--359.

\bibitem{peng2017secret}
Y.~Peng, P.~Wang, W.~Xiang, and Y.~Li, ``Secret key generation based on
  estimated channel state information for tdd-ofdm systems over fading
  channels,'' \emph{{IEEE} Trans. Wireless Commun.}, 2017.

\bibitem{zeng2010exploiting}
K.~Zeng, D.~Wu, A.~Chan, and P.~Mohapatra, ``Exploiting multiple-antenna
  diversity for shared secret key generation in wireless networks,'' in
  \emph{Proc. 29th IEEE Int. Conf. Comput. Commun. (INFOCOM)}, San Diego,
  California, USA, March 2010, pp. 1--9.

\bibitem{wallace2010automatic}
J.~W. Wallace and R.~K. Sharma, ``{Automatic secret keys from reciprocal MIMO
  wireless channels: Measurement and analysis},'' \emph{{IEEE} Trans. Inf.
  Forensics Security}, vol.~5, no.~3, pp. 381--392, 2010.

\bibitem{chen2011secret}
C.~Chen and M.~A. Jensen, ``Secret key establishment using temporally and
  spatially correlated wireless channel coefficients,'' \emph{{IEEE} Trans.
  Mobile Comput.}, vol.~10, no.~2, pp. 205--215, 2011.

\bibitem{jorswieck2013secret}
E.~A. Jorswieck, A.~Wolf, and S.~Engelmann, ``{Secret key generation from
  reciprocal spatially correlated {MIMO} channels},'' in \emph{Proc. IEEE
  GLOBECOM Workshop Trusted Commun. with Physical Layer Security (TCPLS)},
  Atlanta, Georgia, USA, December 2013, pp. 1245--1250.

\bibitem{patwari2010high}
N.~Patwari, J.~Croft, S.~Jana, and S.~K. Kasera, ``High-rate uncorrelated bit
  extraction for shared secret key generation from channel measurements,''
  \emph{{IEEE} Trans. Mobile Comput.}, vol.~9, no.~1, pp. 17--30, January 2010.

\bibitem{zenger2015security}
C.~T. Zenger, J.~Zimmer, and C.~Paar, ``Security analysis of quantization
  schemes for channel-based key extraction,'' in \emph{Proc. 12th EAI Int.
  Conf. Mobile and Ubiquitous Systems: Computing, Networking and Services},
  Coimbra, Portugal, July 2015, pp. 267--272.

\bibitem{dodis2008fuzzy}
Y.~Dodis, R.~Ostrovsky, L.~Reyzin, and A.~Smith, ``Fuzzy extractors: How to
  generate strong keys from biometrics and other noisy data,'' \emph{SIAM J.
  Comput.}, vol.~38, no.~1, pp. 97--139, 2008.

\bibitem{huth2016information}
C.~Huth, R.~Guillaume, T.~Strohm, P.~Duplys, I.~A. Samuel, and T.~G{\"u}neysu,
  ``Information reconciliation schemes in physical-layer security: A survey,''
  \emph{Computer Networks}, vol. 109, pp. 84--104, 2016.

\bibitem{Ali2014Eliminating}
S.~Ali, V.~Sivaraman, and D.~Ostry, ``Eliminating reconciliation cost in secret
  key generation for body-worn health monitoring devices,'' \emph{{IEEE} Trans.
  Mobile Comput.}, vol.~13, no.~12, pp. 2763--2776, December 2014.

\bibitem{premnath2014secret}
S.~N. Premnath, P.~L. Gowda, S.~K. Kasera, N.~Patwari, and R.~Ricci, ``Secret
  key extraction using {Bluetooth} wireless signal strength measurements,'' in
  \emph{Proc. 11th Annu. IEEE Int. Conf. Sensing, Commun. Networking (SECON)},
  Singapore, June 2014, pp. 293--301.

\bibitem{huo2015xor}
F.~Huo and G.~Gong, ``{XOR encryption versus phase encryption, An in-depth
  analysis},'' \emph{{IEEE} Trans. Electromagn. Compat.}, vol.~57, no.~4, pp.
  903--911, August 2015.

\bibitem{reilly2009noise}
D.~Reilly and G.~Kanter, ``Noise-enhanced encryption for physical layer
  security in an {OFDM} radio,'' in \emph{Proc. IEEE Radio and Wireless Symp.
  (RWS)}, San Diego, CA, USA, January 2009, pp. 344--347.

\bibitem{ma2010secure}
R.~Ma, L.~Dai, Z.~Wang, and J.~Wang, ``Secure communication in {TDS-OFDM}
  system using constellation rotation and noise insertion,'' \emph{{IEEE}
  Trans. Consum. Electron.}, vol.~56, no.~3, pp. 1328--1332, 2010.

\bibitem{khan2007secure}
M.~A. Khan, M.~Asim, V.~Jeoti, and R.~S. Manzoor, ``On secure {OFDM} system:
  Chaos based constellation scrambling,'' in \emph{Proc. Int. Conf. on
  Intelligent and Advanced Syst. (ICIAS)}, Kuala Lumpur, Malaysia, November
  2007, pp. 484--488.

\bibitem{tseng2007ofdm}
D.~Tseng and J.~Chiu, ``An {OFDM} speech scrambler without residual
  intelligibility,'' in \emph{Proc. IEEE Region 10 Conf. (TENCON)}, Taipei,
  October 2007, pp. 1--4.

\bibitem{zhang2011secure}
L.~Zhang, X.~Xin, B.~Liu, and Y.~Wang, ``Secure {OFDM-PON} based on chaos
  scrambling,'' \emph{{IEEE} Photon. Technol. Lett.}, vol.~23, no.~14, pp.
  998--1000, 2011.

\bibitem{li2013secure}
H.~Li, X.~Wang, and W.~Hou, ``Secure transmission in {OFDM} systems by using
  time domain scrambling,'' in \emph{Proc. 77th IEEE Veh. Technology Conf. (VTC
  Spring)}, Dresden, Germany, June 2013, pp. 1--5.

\bibitem{li2014dynamic}
H.~Li, X.~Wang, and Y.~Zou, ``Dynamic subcarrier coordinate interleaving for
  eavesdropping prevention in {OFDM} systems,'' \emph{{IEEE} Commun. Lett.},
  vol.~18, no.~6, pp. 1059--1062, June 2014.

\bibitem{li2015eavesdropping}
H.~Li, X.~Wang, and J.-Y. Chouinard, ``Eavesdropping-resilient {OFDM} system
  using sorted subcarrier interleaving,'' \emph{{IEEE} Trans. Wireless
  Commun.}, vol.~14, no.~2, pp. 1155--1165, February 2015.

\bibitem{zhang2016design}
J.~Zhang, A.~Marshall, R.~Woods, and T.~Q. Duong, ``Design of an {OFDM}
  physical layer encryption scheme,'' \emph{{IEEE} Trans. Veh. Technol.},
  vol.~66, no.~3, pp. 2114--2127, 2017.

\bibitem{nain2017secure}
A.~K. Nain, J.~Bandaru, M.~A. Zubair, and R.~Pachamuthu, ``A secure
  phase-encrypted ieee 802.15.4 transceiver design,'' \emph{IEEE Trans.
  Computers}, vol.~66, no.~8, pp. 1421 -- 1427, August 2017.

\bibitem{wan2016exploiting}
J.~Wan, A.~B. Lopez, and M.~A. Al~Faruque, ``Exploiting wireless channel
  randomness to generate keys for automotive cyber-physical system security,''
  in \emph{Proc. 7th International Conference on Cyber-Physical Systems},
  Vienna, Austria, April 2016, p.~13.

\bibitem{zhu2017using}
X.~Zhu, F.~Xu, E.~Novak, C.~C. Tan, Q.~Li, and G.~Chen, ``Using wireless link
  dynamics to extract a secret key in vehicular scenarios,'' \emph{IEEE Trans.
  Mobile Comput.}, vol.~16, no.~7, pp. 2065--2078, April 2017.

\bibitem{li2017efficient}
X.~Li, J.~Liu, Q.~Yao, and J.~Ma, ``Efficient and consistent key extraction
  based on received signal strength for vehicular ad hoc networks,'' \emph{IEEE
  Access}, vol.~5, pp. 5281--5291, 2017.

\bibitem{abdelgader2014secret}
A.~M. Abdelgader and L.~Wu, ``A secret key extraction technique applied in
  vehicular networks,'' in \emph{Proc. IEEE 17th Int. Conf. Computational
  Science and Engineering}, Chengdu, China, December 2014, pp. 1396--1403.

\bibitem{abdelgader2017exploiting}
A.~M.~S. Abdelgader, S.~Feng, and L.~Wu, ``Exploiting the randomness inherent
  of the channel for secret key sharing in vehicular communications,''
  \emph{International Journal of Intelligent Transportation Systems Research},
  pp. 1--12, 2017.

\bibitem{vogt2016practical}
H.~Vogt, K.~Ramm, and A.~Sezgin, ``Practical secret-key generation by
  full-duplex nodes with residual self-interference,'' in \emph{Proc. 20th Int.
  ITG Workshop on Smart Antennas}, Munich, Germany, March 2016, pp. 344--347.

\bibitem{wang2011fast}
Q.~Wang, H.~Su, K.~Ren, and K.~Kim, ``Fast and scalable secret key generation
  exploiting channel phase randomness in wireless networks,'' in \emph{Proc.
  30th IEEE Int. Conf. Comput. Commun. (INFOCOM)}, Shanghai, China, April 2011,
  pp. 1422--1430.

\bibitem{liu2014group}
H.~Liu, J.~Yang, Y.~Wang, Y.~J. Chen, and C.~E. Koksal, ``Group secret key
  generation via received signal strength: Protocols, achievable rates, and
  implementation,'' \emph{{IEEE} Trans. Mobile Comput.}, vol.~13, no.~12, pp.
  2820--2835, 2014.

\bibitem{thai2015secret}
C.~D.~T. Thai, J.~Lee, and T.~Q. Quek, ``Secret group key generation in
  physical layer for mesh topology,'' in \emph{Proc. IEEE Global Communications
  Conference (GLOBECOM)}, San Diego, CA, USA, December 2015, pp. 1--6.

\bibitem{zenger2016passive}
C.~Zenger, H.~Vogt, J.~Zimmer, A.~Sezgin, and C.~Paar, ``The passive
  eavesdropper affects my channel: Secret-key rates under real-world
  conditions,'' in \emph{Proc. IEEE GLOBECOM Workshop Trusted Commun. with
  Physical Layer Security (TCPLS)}, Washington DC, USA, December 2016, pp.
  1--6.

\bibitem{zafer2012limitations}
M.~Zafer, D.~Agrawal, and M.~Srivatsa, ``Limitations of generating a secret key
  using wireless fading under active adversary,'' \emph{{IEEE/ACM} Trans.
  Netw.}, vol.~20, no.~5, pp. 1440--1451, October 2012.

\bibitem{jin2015physical}
R.~Jin and K.~Zeng, ``Physical layer key agreement under signal injection
  attacks,'' in \emph{Proc. IEEE Conf. Commun. and Network Security (CNS)},
  Florence, Italy, September 2015, pp. 254--262.

\end{thebibliography}

\end{document}